\documentclass[lettersize,journal]{IEEEtran}
\usepackage{amssymb}
\usepackage{bbm}
\usepackage[linesnumbered,ruled,vlined]{algorithm2e}
\usepackage{bibentry}
\usepackage{amsfonts}
\usepackage{subfigure}
\usepackage{multirow}
\usepackage{amsmath}
\usepackage{graphicx}
\usepackage{epsfig}
\usepackage{color}
\usepackage{epstopdf}
\usepackage{rotating}
\usepackage{caption}
\usepackage{float}
\usepackage{multicol}
\usepackage{amssymb}
\usepackage{graphicx}
\usepackage{bbding}
\usepackage{balance}
\usepackage{microtype}
\usepackage{graphicx}
\usepackage{subfigure}
\usepackage{booktabs} % for professional tables
\usepackage{multirow}
\usepackage{amsmath}
\usepackage{amsmath, bm}

\usepackage{mathtools}
\usepackage{etoolbox}
\usepackage{cases}
\usepackage{enumitem}
\usepackage{xcolor}
\usepackage{subfigure}
\usepackage{xcolor}
\usepackage{color}
\usepackage{amssymb}

\usepackage{cases}

\usepackage{cite}
\usepackage{amsmath,amssymb,amsfonts}
\usepackage{graphicx}
\usepackage{textcomp}

\usepackage{mathtools}
\usepackage{amssymb}
\usepackage{amsthm}
\usepackage{color}
\usepackage{multirow}
\usepackage{amssymb}

\usepackage{cases}
\usepackage{cite}
\usepackage{amsmath,amssymb,amsfonts}
\usepackage{graphicx}
\usepackage{textcomp}
\usepackage{xcolor}
\definecolor{brown}{RGB}{139,64,0}
\definecolor{pink}{RGB}{255,170,182}
\definecolor{purple}{RGB}{160,32,240}
\usepackage{amssymb}

\usepackage{cases}

\usepackage{cite}
\usepackage{amsmath,amssymb,amsfonts}
\usepackage{graphicx}
\usepackage{textcomp}
\usepackage{xcolor}

\begin{document}

\title{Joint Universal Adversarial Perturbations \\ with Interpretations}

\author{Liang-bo Ning$^\dagger$, Zeyu Dai$^\dagger$, Wenqi Fan$^\ddagger$, Jingran Su, Chao Pan, \\ Luning Wang, Qing Li$^\ddagger$,~\IEEEmembership{Fellow,~IEEE}
        % <-this % stops a space

\thanks{$^\dagger$Equal contributions: Liang-bo Ning and Zeyu Dai; $^\ddagger$Corresponding authors: Dr. Wenqi Fan and Prof.  Qing Li.}
\thanks{Liang-bo Ning, Zeyu Dai, Jingran Su, Chao Pan, Qing Li are with the Department of Computing, The Hong
Kong Polytechnic University, Hong Kong. (E-mail: BigLemon1123@gmail.com; ze-yu.dai@connect.polyu.hk; jing-ran.su@connect.polyu.hk; chao.pan@connect.polyu.hk; qing-prof.li@polyu.edu.hk.) }% <-this % stops a space
\thanks{Wenqi Fan is with the Department of Computing and the Department of Management and Marketing, The Hong Kong Polytechnic University, Hong Kong. (E-mail: wenqifan03@gmail.com.) }
\thanks{Luning Wang is with Huawei Hong kong Research Center, Hong Kong. (E-mail: wangluning2@huawei.com.) }
}

\markboth{Journal of \LaTeX\ Class Files,~Vol.~14, No.~8, August~2021}%
{Shell \MakeLowercase{\textit{et al.}}: A Sample Article Using IEEEtran.cls for IEEE Journals}


\maketitle

\begin{abstract}
Deep neural networks (DNNs) have significantly boosted the performance of many challenging tasks. Despite the great development, DNNs have also exposed their vulnerability. Recent studies have shown that adversaries can manipulate the predictions of DNNs by adding a universal adversarial perturbation (UAP) to benign samples. On the other hand, increasing efforts have been made to help users understand and explain the inner working of DNNs by highlighting the most informative parts (i.e., attribution maps) of samples with respect to their predictions. Moreover, we first empirically find that such attribution maps between benign and adversarial examples have a significant discrepancy, which has the potential to detect universal adversarial perturbations for defending against adversarial attacks. This finding motivates us to further investigate a new research problem:  whether there exist universal adversarial perturbations that are able to jointly attack DNNs classifier and its interpretation with malicious desires. It is challenging to give an explicit answer since these two objectives are seemingly conflicting. In this paper, we propose a novel attacking framework to generate joint universal adversarial perturbations (\textbf{JUAP}), which can fool the DNNs model and misguide the inspection from interpreters simultaneously. Comprehensive experiments on various datasets demonstrate the effectiveness of the proposed method JUAP for joint attacks. To the best of our knowledge, this is the first effort to study UAP for jointly attacking both DNNs and interpretations.

\end{abstract}

\begin{IEEEkeywords}
Adversarial Attacks, Universal Adversarial Perturbations (UAP), Interpretation, Trustworthy AI.
\end{IEEEkeywords}

\section{Introduction}

As one of the most important data mining techniques, Deep Neural Networks (DNNs) have achieved massive remarkable performance in various challenging downstream domains~\cite{XAI,fan2022comprehensive,ding2024survey,liu2022new}, e.g., image classifications~\cite{zhao2024evaluating, xu2023channel,zhang2023representation,zhang2023mining},  language generation ~\cite{yao2024survey,pan2024unifying}, and social media~\cite{ding2019social,wang2023multi}. 
These deep learning methods have significantly benefited our humans in daily life and created great economic outcomes in society. 
Along with their impressive development and great achievement, DNNs techniques have also exposed their untrustworthy aspects~\cite{li2024bic,li2023trustworthy,fan2023adversarial,wei2024trustworthy,ning2024interpretation}, which might perform unreliable predictions and cause severe economic, social, and security consequences, especially in safety-critical scenarios.
For example, adversaries can maliciously generate well-designed patterns in road signs, such that DNNs-enhanced autonomous vehicles might recognize a stop sign as an acceleration sign \cite{stallkamp2012man}, leading to great potential risks to personal safety.

Current adversarial perturbations in computer vision tasks can be divided into two types,
i.e., image-dependent perturbations (\textbf{IDPs}) and image-agnostic universal adversarial perturbations~(\textbf{UAPs}). 
Image-dependent perturbations need to be crafted for each image by using iterative/non-iterative gradient decent~\cite{pgd,fgsm,bortolussi2024robustness,baniecki2024adversarial,li2024survey} algorithms or solving an optimization problem~\cite{bastani2016measuring}. Instead, image-agnostic UAP can be added to most benign images and then cause misclassification of DNNs with very high confidence~\cite{uap,pan2024adversarial}. Due to the good generalization capability, UAPs, after learning from the dataset, can even attack unseen images. Besides, compared with image-dependent perturbations specific to each sample, image-agnostic UAPs, once learned and deployed, are more efficient, especially when facing large-scale data samples. Thus,  image-agnostic universal adversarial perturbations pose a significant threat to the security of DNNs.

\begin{figure}[t]
% \vskip -0.2in
\centering
\includegraphics[width=0.98\columnwidth]{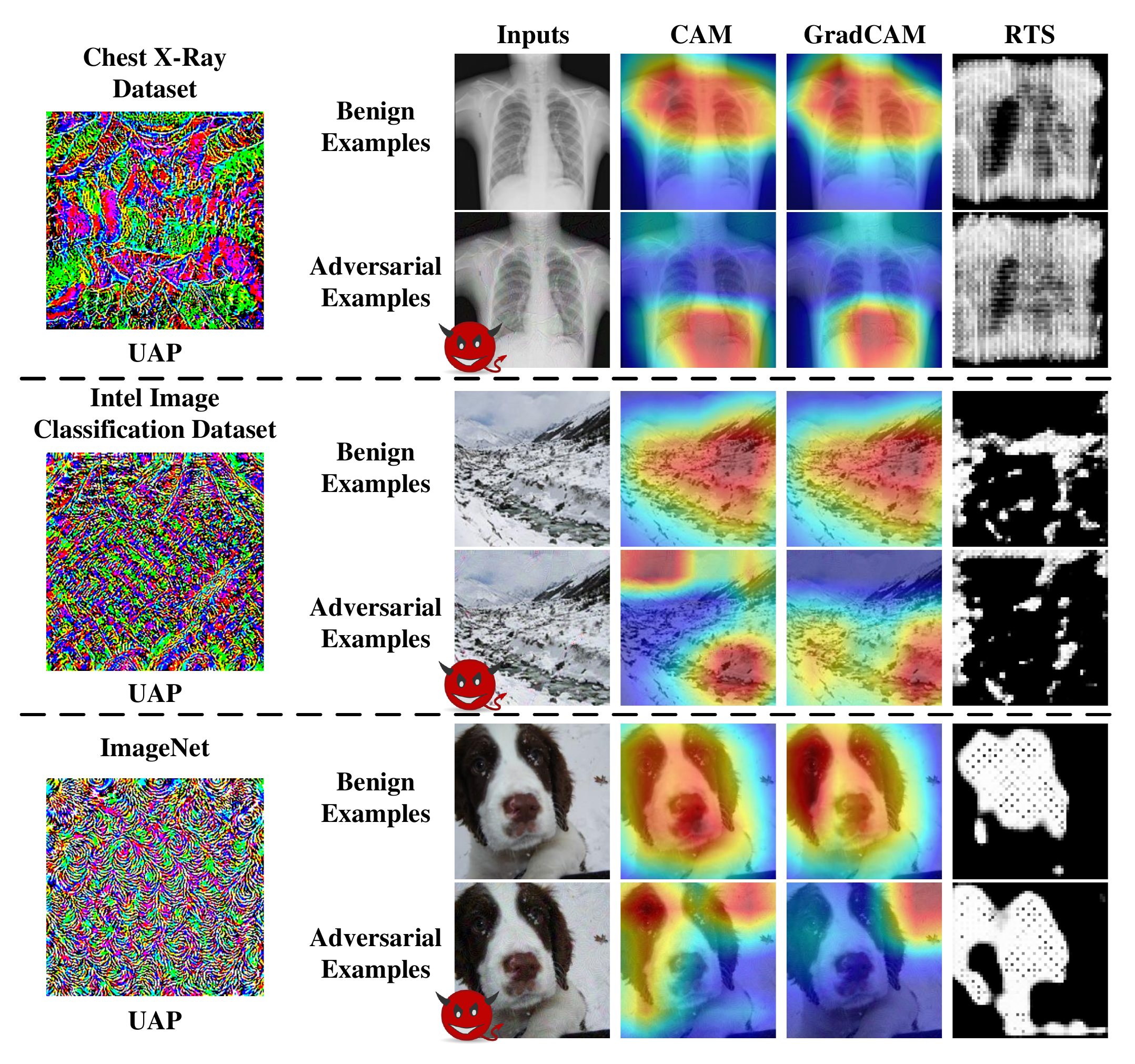}
% \vskip -0.180in
\caption{An illustration of interpretation discrepancy for benign and adversarial images (via adding universal adversarial perturbations) under various interpreters.}
\label{JUAP}
 % \vskip -0.2in
\end{figure}

In recent years, in order to enhance the trustworthiness of DNNs models, increasing efforts have been made to help users understand the inner working mechanism of DNNs models by providing interpretations on how certain decisions are made~\cite{cam, gradcam, he2023harnessing,novakovsky2023obtaining}.
More specifically, the interpretations can be achieved by identifying the most influential parts (e.g., attribution maps) of the input with respect to its prediction~\cite{grad, rts, lime, mask, sharp,dwivedi2023explainable}.
For example, as the first activation-based interpreter, CAM was proposed to compute a weighted sum of the activation of the last convolutional layer to generate attribution maps,  
where the weights are the parameters of the final dense layer~\cite{cam}. 
Simonyan et al.~\cite{grad} proposed to use the derivative as the attribution maps, considered as the gradient-based method. 
On the other hand, recent works have found that DNNs models' interpretability can work as inspectors or detectors (i.e., model debugging) to diagnose models' errors and reveal models' vulnerability (i.e., adversarial perturbations) in computer vision domain~\cite{adebayo2020debugging} and text domain~\cite{lertvittayakumjorn2021explanation,lertvittayakumjorn2020find,ribeiro2018semantically}.
As a result, interpreters have the potential to detect adversarial perturbations for defending against adversarial attacks.

In order to verify such detection capabilities in the image domain under various  DNNs' interpreters (i.e., CAM, GradCAM, and RTS) for UAP attacks~\cite{uap}, we first perform empirical studies to evaluate the interpretation discrepancy between interpretation maps on the benign and adversarial images.   
As shown in Figure~\ref{JUAP}, the attribution maps on adversarial examples are significantly different from those of benign examples. 
Once such `confused' predictions are inspected, humans (e.g., model designers) can develop some strategies to enhance the DNNs model's robustness against adversarial attacks.
In other words, interpreters can be further exploited as anomaly detectors to improve the system's safety.
Thus, our experimental results validate that \emph{DNNs' interpretations provide great opportunities to detect universal adversarial perturbations for defending against adversarial attacks.}

In order to evade interpreters' detection with malicious desires, we further raise a new question, i.e., 
\emph{whether there exist universal adversarial perturbations that can perform a joint attack to attack DNNs classifier and its interpretation simultaneously with malicious desires?} 
To be specific, by adding the universal adversarial perturbations to the benign images, the objectives of the joint attack are to maximize the change of DNNs' predictions while minimizing the interpretation discrepancy, leading to more severe safety issues for DNNs techniques.
However, these two objectives are essentially contradictory to each other. That is because the added universal perturbations tend to be the most influential part of natural images for obtaining incorrect labels, while the goal of interpreters is to identify the most influential maps for explaining how do DNNs models make decisions. 
Therefore, to investigate such new questions, in this paper, we propose a novel attacking framework (\textbf{JUAP}) for generating universal adversarial perturbations, which can simultaneously mislead the DNNs model's prediction and misguide the inspection from the interpreter.  
%More specifically,  \wq{an optimization}

Our major contributions  are summarised as follows:
%\vspace{-1mm}
\begin{itemize}%[leftmargin=*] 
    \item We discover that  DNNs modes’ interpretability can be leveraged to understand and inspect the problematic outputs from adversarial examples caused by universal adversarial perturbations, which pave the way to enhance the safety of DNNs models.
    
    \item We study a novel problem of exploring universal adversarial perturbations to jointly attack DNNs models as well as their interpretations, which reveals more severe safety concerns.
    To the best of our knowledge, we are the first to investigate universal adversarial perturbations from the perspective of interpreters' detection.

    \item We propose a novel attacking framework to generate joint universal adversarial perturbations (\textbf{JUAP}), where a generative adversarial network is proposed to learn universal perturbations for fooling the DNNs model and misguide the inspection from interpreters simultaneously. 

    \item  Our comprehensive experiments on three real-world datasets demonstrate that the proposed attacker can achieve a high fooling ratio for attacking classifiers with a small interpretation discrepancy, which successfully achieves the joint attack on DNNs and their coupled interpreters.
\end{itemize}

The remainder of this paper is organized as follows.  
In Section~\ref{sec:problem_definition}, we introduce the problem definition. Then, we present the details of our proposed method to generate joint universal adversarial perturbations in Section \ref{sec:methodology}. 
In Section \ref{sec:experiments}, we conduct massive experiments and demonstrate the effectiveness of our JUAP. 
In Section \ref{sec:rw}, we briefly review some related works. 
At last, we conclude our work with future directions in Section~\ref{sec:conclusion}.

\section{Problem Definition}~\label{sec:problem_definition}
 
Given a benign image $x\in \mathbb{R}^d$ with its ground truth $C_x \in \{1,2,...,k\}$, the goal of a DNN classifier is to learn a model for label prediction, which can be formulated as $f(x): \mathbb{R}^d\rightarrow \{1,2,...,k\}$.

\noindent \textbf{Adversarial Perturbations}.
For a benign image $x$,  attackers can generate imperceptible perturbations $\hat{n}$ to obtain its corresponding adversarial example $\hat{x} = x + \hat{n}$  so as to manipulate the DNN classifier $f(\hat{x})$ to make the malicious prediction. 
Moreover, the perturbations are required to satisfy the imperceptibility property (i.e., perceptually indistinguishable from our human eye), which can be achieved by adding constraint: ${\left\| {\hat n} \right\|}_p \le \zeta$,
where $\zeta$ is a hyperparameter used to control the magnitude of the perturbation and $p$ is the vector norm. 

In general, adversarial perturbations can be categorized as image-dependent and universal (image-agnostic)  perturbations.
Given a well-trained DNNs classifier $f$ and a dataset $D=\{x_i,C_{x_i}|i=1,...,N\} $ containing $N$ samples with their labels~$C_{x_i}$:
 \begin{itemize} %[leftmargin=*] 
     \item \textbf{Image-dependent Adversarial Perturbations} can be defined as ${\hat n}_i$ for a specific image $x_i$. 
     The image-dependent perturbations can mislead $f(x_i)$ to provide wrong prediction output, which can be mathematically formulated as: 
     \begin{equation}
     f(x_i) \ne f(x_i + \hat n_i), i=1,\cdots, N.
    \end{equation}
     
     Normally, image-dependent perturbations are devised for each sample, respectively, which is time-consuming and lacks generalization ability. 
     The perturbations generated based on one specific sample may not function well when added to other benign images.

     \item \textbf{Universal Adversarial Perturbations (UAPs)} are defined as $\hat{n}$, aiming to fool classifer $f$ to produce incorrect predictions for most samples in $D$ as follows: 
     \begin{equation}
     f(x_i) \ne f(x_i + \hat n), i=1,\cdots, N.
     \end{equation}
     
     The universal adversarial perturbations are usually learned based on a set of training samples with great generalization properties. Thus, UAPs can also mislead the DNNs' predictions when added to unseen benign samples.

 \end{itemize}

\noindent \textbf{DNNs' Interpretations}. 
Given a classifier $f$, its coupled interpreters are defined as $\mathcal{I}: \mathbb{R}^d\rightarrow \mathbb{R}^d$, aiming to produce an attribution map $\mathcal{I} (x)$, which highlights the most influential parts of the input with respect to its prediction. 
The attribution map can provide a great way to help users understand and explain the inner working mechanism of how certain decisions are made from the DNNs model.
Moreover, recent works have found that  DNNs' interpreters have the potential to work as anomaly detectors for improving the security of DNNs against adversarial attacks~\cite{adebayo2020debugging,lertvittayakumjorn2021explanation,lertvittayakumjorn2020find,ribeiro2018semantically}, as demonstrated in our empirical studies (see the Figure~\ref{JUAP}).

\noindent \textbf{Problem Statement: }
Given an image $x_i$, the attacker aims to perform a joint attack by generating universal adversarial perturbations $\hat{n}$,  which can fulfill the following two goals:
\begin{itemize}%[leftmargin=*] 
\item  The UAP $\hat{n}$ on any sample $x_i$  can change the DNNs classifier prediction to incorrect label:
\begin{equation}
f(x_i) \ne f(x_i + \hat n), i=1, \cdots, N;
\end{equation}

\item The UAP $\hat{n}$ on any sample $x_i$  can manipulate the interpreter to produce the same attribution maps as that on that original input: 
\begin{equation}
\mathcal{I}(x_i)=\mathcal{I}(x_i+\hat n), i=1, \cdots,N. 
\end{equation}

\end{itemize}

\section{Methodology}\label{sec:methodology}
In this section, we introduce the details of our proposed method \textbf{JUAP} to achieve the joint attack.

\begin{figure*}[ht]
%\setlength{\abovecaptionskip}{3mm}
% \vskip -0.1in
\centering
\includegraphics[width=1\linewidth]{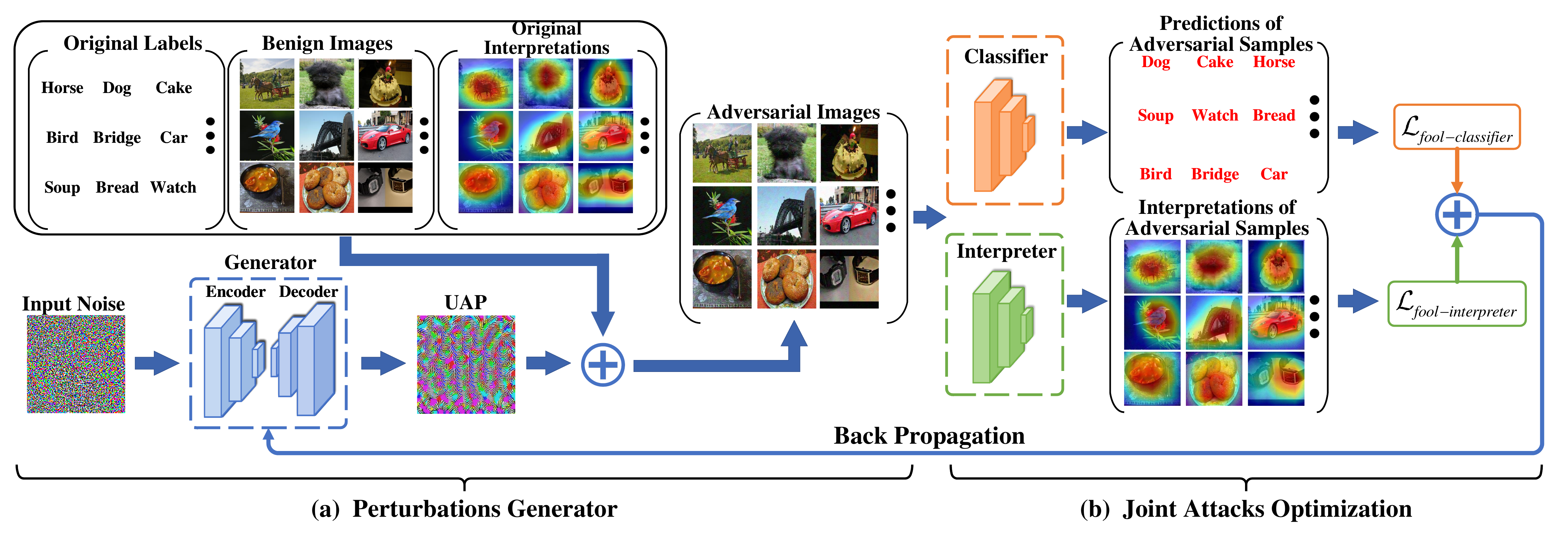}
% \vskip -0.1in
\caption{The overall framework of the proposed method (JUAP), consisting of perturbations generator and joint attacks optimization for fooling DNNs classifier and interpreter simultaneously.}
% \vskip -0.17in
\label{JUAP_framework}
\end{figure*}

\subsection{An Overview of Our Proposed Method}
The goal of this work is to conduct joint attacks by learning image-agnostic adversarial perturbations. 
To achieve this goal, we propose a novel universal adversarial perturbations framework (\textbf{JUAP}) to jointly attack the DNNs classifier and  interpreter. 
More specifically, a generative adversarial network is proposed to learn universal perturbations that cause natural samples to be misclassified with high confidence while keeping their attribution maps unchanged.  
The overall framework of the proposed method JUAP is shown in Figure~\ref{JUAP_framework}, which consists of perturbations generator and joint attacks optimization for fooling DNNs classifier and  interpreter simultaneously.

\subsection{Attack DNNs Classifiers}\label{methodology_JUAP}
A general solution to generate universal adversarial perturbations is to leverage an iterative  gradient-based strategy~\cite{uap, fastuap}. 
However, iteratively creating perturbations for each sample and fusing all perturbations to get the UAP is time-consuming.
Moreover, UAP created by data-driven iterative gradient-based methods significantly depends on the training set. 
Thus, in this work, in order to attack DNNs classifiers, a generative adversarial framework is proposed to learn universal adversarial perturbations $\hat{n}$, which can be formulated as an image-to-image translation task by reconstructing different noise vectors, consisting of an encoder $e(\cdot)$ and a decoder $d(\cdot)$.
Given various initial input noise, the output perturbations are different from each other. 
Meanwhile, this generative model can generate infinite UAPs based on a specific training set. 
To accelerate the convergence speed, we randomly sample a noise vector from the uniform distribution $U(0, 1)$, denoted as $n$, with the same size as the input images. 
More specifically, a generator $\mathcal{G_{\theta}}$ with weights $\theta$ is developed to map the random noise image to joint universal adversarial perturbations $\hat n$, which can be modeled as follows:
\begin{equation}
\hat{n} =  \mathcal{G_{\theta}}(n) = d(e(n)), n \sim {U}(0, 1), 
\end{equation}
\noindent where the encoder extracts the latent variables (or semantic features) of the inputs, while the decoder receives the encoders' output and reconstructs the latent variables to implement the image-to-image translation. 
More details about generator can be found in Section~\ref{sec:UAP-G}.

To make the JUAP imperceptible, the output of the generator is scaled to have a fixed magnitude.
Mathematically, the JUAP generation process can be defined as follows:
\begin{equation}
\hat n = \min \left( {1,\frac{\zeta }{{{{\left\| {{\mathcal{G}_\theta }\left( n \right)} \right\|}_p}}}} \right) \cdot {\mathcal{G}_\theta }\left( n \right), n \sim {U}(0, 1),
    \label{generator}
\end{equation}
where $\zeta$ is a pre-defined threshold to control the magnitude of the JUAP. ${\left\| {\cdot} \right\|}_p$ is the $p$-norm of the vector. 
After that, the adversarial examples can be obtained by adding the joint universal perturbation to benign images, i.e., $\hat x = x + \hat n$.

To optimize the parameters $\theta$ of the generator $\mathcal{G}$ for attacking DNNs classifiers, we need to specify a classification objective to optimize.
More specifically, with different attacking goals, we can define three types of objectives to learn generator $\mathcal{G_{\theta}}$'s parameters as follows: 
\begin{itemize} % [leftmargin=*] 
\item 
One of the representative attacking goals is to perform \emph{non-targeted attack}. 
The most general practice for the image classification task
is to minimize the cross-entropy loss between predictions and true labels to guide the parameters' optimization for the exact classification.
Inspired by this paradigm, we propose to maximize the cross-entropy loss for the non-targeted attack as follows:
\begin{equation} 
\min_\theta{{{\mathcal{L}}}_{fool - classifier}} =  \min_\theta ~ - \log ( {{\mathcal{L}}}\left( {f\left( {\hat x} \right),{C_x}} \right) )
\label{cross},
\end{equation}
where $\mathcal{L}$ is the cross-entropy loss.

\item 
Instead of maximizing the cross-entropy loss, other alternatives can also be exploited to deceive classifiers for \emph{non-targeted attack}. 
The most straightforward method is to assign a wrong label for samples at each iteration and minimize the cross-entropy loss to attack classifiers. To achieve a great performance, we set the least likely class as the target label as follows: 
\begin{equation}
\min_\theta{\mathcal{{L}}_{fool - classifier}} = \min_\theta ~\log \left( {\mathcal{{L}}\left( {f\left( {\hat x} \right),{C_{\min }}} \right)} \right),
\label{cmin}
\end{equation}
where $C_{min}$ is the class with the minimum prediction probability, i.e., $C_{min} = \arg \min(f(x))$.

\item Next, we can consider a more difficult task:  \emph{targeted attack}. 
In such a scenario, the goal of universal perturbations is to not only mislead the classifiers but also change the predictions to a specific target label $C_{t}$. To achieve the targeted attack, we can directly minimize the cross-entropy loss toward a target label $C_{t}$:
\begin{equation}
\min_\theta{\mathcal{{L}}_{fool - classifier}} = \min_\theta ~\log \left( {\mathcal{{L}}\left( {f\left( {\hat x} \right),{C_{t}}} \right)} \right)
\label{target}.
\end{equation}

\end{itemize}

\subsection{Attack DNNs Interpreters}
Fooling interpreters aims to minimize the discrepancy of interpretation maps between benign and adversarial examples. 
To demonstrate the effectiveness of the proposed methods, 
we focus on adversarial attacks on three representative interpretation methods, i.e., CAM~\cite{cam}, GradCAM~\cite{gradcam}, and RTS~\cite{rts}, where they have different underlying mechanisms. 
CAM~\cite{cam} is the representative activation-based interpreter, which has been widely used in various computer vision tasks. Using GradCAM~\cite{gradcam} as the interpreter aims to demonstrate the effectiveness of JUAP in manipulating attribution maps based on gradients. Here typical gradient-based interpreters~\cite{Smoothgrad} are not taken into account since the attribution maps produced by these methods are not sufficiently clear/distinguishable. RTS~\cite{rts} is employed as a representation of methods that do not rely on the intrinsic model parameters (e.g., activations or gradients) for generating the attribution maps.
Mathematically,   the aforementioned interpreter methods are briefly summarised as follows:
\begin{itemize} %[leftmargin=*] 
\item CAM is the first activation-based method, which is easily embedded into most existing DNNs. Let $A^k_{ij}$ denote the element in the $i$-th row and $j$-th column of the $k$-th channel of the last convolutional layer. The output of the following global average pooling layer is ${A^k} = \frac{1}{V}\sum\nolimits_i {\sum\nolimits_j {A_{ij}^k} }$, where $V$ is the number of the latent variables in the last convolutional layer.
$\omega^k_c$ denotes the weight in final dense layer between $A^k$ and the logits of class $c$. The attribution map for class $c$ is defined as $\mathcal{I}_c$, where each spatial element is given by
\begin{equation}
    \mathcal{I}_c(i,j) = \sum_k \omega_c^k A_{ij}^k.
    \label{cam}
\end{equation}

\item 
GradCAM is one of the variants of CAM, which takes the gradient into consideration. The gradient is used to compute the weight $\omega^k_c$, i.e.,
\begin{equation}
    \omega _c^k = \frac{1}{V}\sum\nolimits_i {\sum\nolimits_j {\frac{{\partial {f_c}}}{{\partial A_{ij}^k}}} },
\end{equation}
where $f_c$ is the $c$-th output of the classifier. The attribution maps are computed by Eq~\eqref{cam}.

\item RTS is the representative model-based method, which trains a model to create attribution maps. Let $R_\varpi$ denote the interpreter with weights $\varpi$. The attribution maps are created by the interpreter, i.e., $\mathcal{I} = {R_\varpi }\left( x \right)$. 
To optimize the parameters of RTS interpreter, the objective function is formulated as:
\begin{align}
\mathcal{L}_{rts}  =& {\lambda _1}{\mathcal{L}_{tv}}(\mathcal{I}) + {\lambda _2}{\mathcal{L}_{av}}(\mathcal{I}) \\
 \nonumber &- \log \left( {{f_c}(\phi (x,\mathcal{I}))} \right) + {\lambda _3}{f_c}{(\phi (x,1 - \mathcal{I}))^{{\lambda _4}}},
\end{align}
where ${\mathcal{L}_{tv}}(\mathcal{I})$ and ${\mathcal{L}_{av}}(\mathcal{I})$ are the total variation and the average value of the attribution map. These two items aim to reduce the noise and retain the most informative part of the attribution map. $\phi$ uses the interpretation $\mathcal{I}$ as the mask and blends the benign image $x$ with noise. The last two items encourage the interpreter to capture the most informative part of the attribution map.
\end{itemize}

In order to deceive DNNs' interpreters, we aim to minimize the interpretation shift of interpretation maps between benign and adversarial examples.
There are multiple alternatives, e.g., Mean Square Error, Kullback-Leibler Divergence, Maximum Mean Discrepancy, etc. In this paper, we use the ${L}_2$ measure, which is easy to implement without additional computation cost. 
The objective is shown as follows:
\begin{equation}
\min_\theta {\mathcal{L}_{fool - interpreter}} = \min_\theta ~ {\left\| {\mathcal{I}\left( {\hat x} \right) - \mathcal{I}\left( x \right)} \right\|_2}~,
\label{foolint}
\end{equation}
\begin{equation}
\mathcal{I}(i,j)\left(  \cdot  \right) = \left\{ \begin{array}{l}
\sum\limits_k {\omega _c^k \cdot A_{ij}^k }, \text{~if~CAM~,} \\
\sum\limits_k {\left( {\frac{1}{V}\sum\nolimits_i {\sum\nolimits_j {\frac{{\partial {f_c}}}{{\partial A_{ij}^k}}} } } \right) \cdot A_{ij}^k}, \text{~if~GradCAM~,} \\
{R_\varpi }\left(  \cdot  \right), \text{~if~RTS~.}
\end{array} \right.
\end{equation}

\noindent \textbf{Optimization}. 
During deceiving interpreters, it is necessary to ensure that the second-order derivative of generators is not all-zero, especially for the interpreters using gradients. The ReLU activation function is widely used in various DNNs, which consists of two linear components. The second-order derivative of ReLU is all-zero and brings difficulties in performing back-propagation during gradient descent. 
To tackle this optimization issue, we propose using a smoothing mechanism on the gradient of ReLU for minimizing Eq.~\eqref{foolint} as follows: 
\begin{equation}
a\left( s \right) = \mathbbm{1}\left( {s < 0} \right) \cdot ( 1 + s/\sqrt {{s^2} + \tau } )  + \mathbbm{1} \left( {s \ge 0} \right) \cdot s/\sqrt {{s^2} + \tau },  
\label{relu}
\end{equation}
where $\mathbbm{1}(\cdot)$ is the indicator function, and $\tau = 10^{-4}$ is a small constant to ensure that the output of the function is bounded. $s$ is the input of the ReLU activation layer.

\subsection{Universal Adversarial Perturbations for Joint Attack}

After introducing the DNNs classifier attack and interpreters attack, we finally design the joint attack to generate UAPs for attacking the classifier and its interpreter simultaneously.
The most straightforward approach for the joint attack is to minimize the aforementioned loss functions simultaneously for optimizing generator's parameters. 
However, deceiving both classifiers and interpreters still face a tremendous challenge. That is, the objective of changing predictions usually contradicts the objective of keeping interpretations unchanged. 
That is because to change the classifier output more significantly, the perturbations tend to be added to the most informative parts in an image, which are also the regions easily detected by the interpreter.

To address the challenge, we propose an iterative optimization strategy for optimizing the objectives of classifier and interpreter attacks.
More specifically, at the beginning of the generator training, we conduct joint optimization on the two objectives. 
When the perturbations produced by the generator are able to deceive the classifier, we focus on fooling its coupled interpreter without deceiving classifiers. 
The overall objective is defined as follows:
\begin{equation}
 {\mathcal{{\cal L}}_{JUAP}} = \max \{ {\mathcal{{\cal L}}_{fool - classifier}} , \delta \} + \lambda \cdot {\mathcal{{\cal L}}_{fool - {interpreter}}},
\label{final}
\end{equation}
where $\lambda$ is a hyperparameter to control the attacking effect between the classifier and interpreter, and $\delta$ is exploited to decide whether to fool classifiers. 
In the following section, we conduct comprehensive experiments and demonstrate that the proposed method is robust to different loss functions defined in Eq.~\eqref{cross}-\eqref{target}.
The overall training process of the proposed method JUAP is shown in Algorithm \ref{al_juap}.

\subsection{JUAP for  Image-dependent Perturbations}

\begin{figure}[t]
\setlength{\abovecaptionskip}{3mm}
% \vskip -0.15in
\centering
\includegraphics[width=0.998\linewidth]{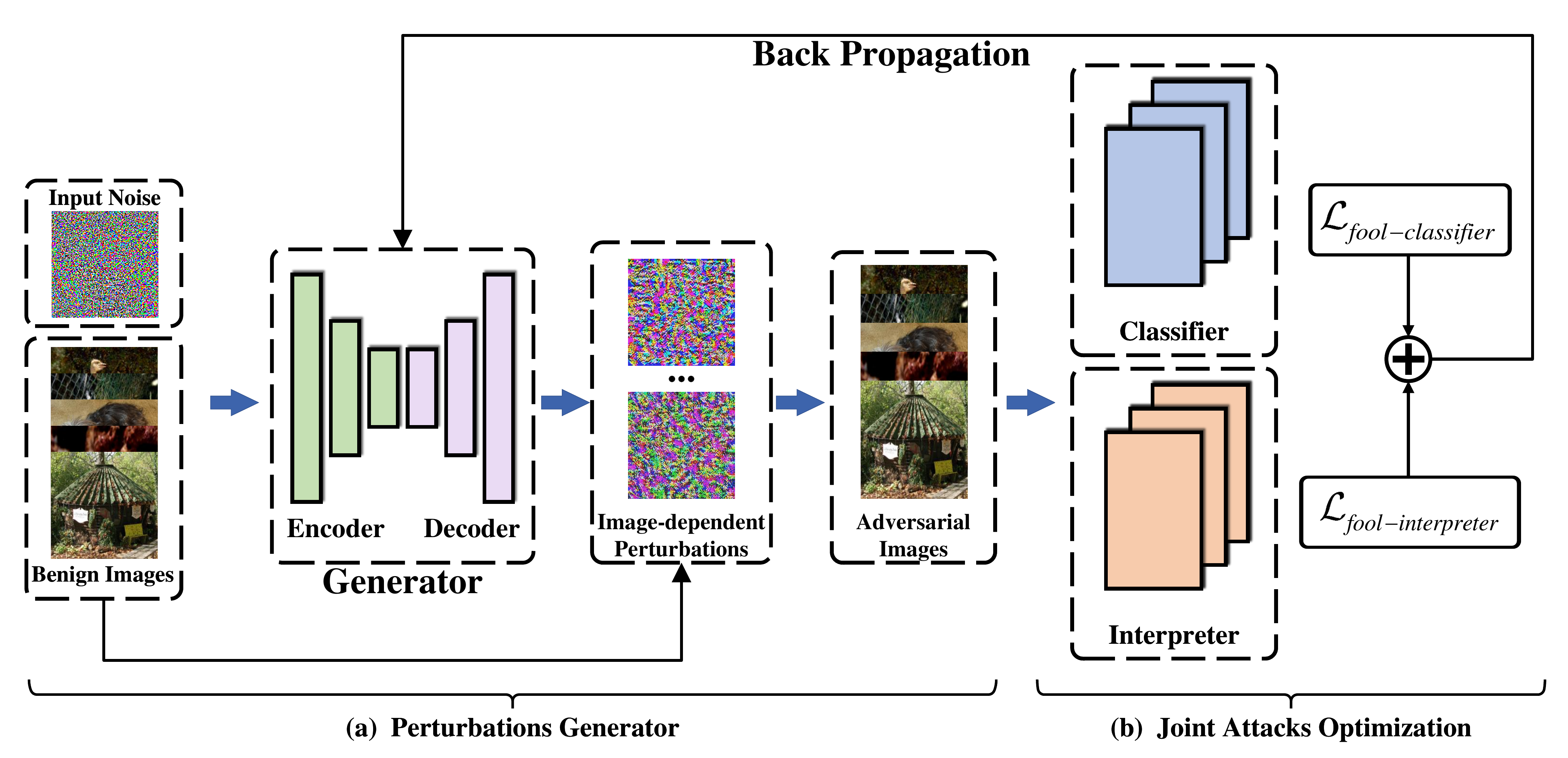}
% \vskip -0.1in
\caption{The overall framework of the proposed method on for generating joint image-dependent perturbations.}
\label{JUAP_individual}
% \vskip -0.1in
\end{figure}

Although our proposed method is designed for generating image-agnostic perturbations,
the proposed framework has strong scalability and can be easily generalized to produce image-dependent perturbations.
To achieve the goal, instead of using multiple samples to learn universal adversarial perturbations, image-dependent perturbations are devised for each sample. 
That is, for each sample, the parameters of the generator vary and are updated by backpropagation for learning image-dependent perturbations.
The production of image-dependent perturbations is defined by:
\begin{equation}
{{\hat n}_i} = \min \left( {1,\frac{\zeta }{{{{\left\| {{\mathcal{{\cal G}}_{{\theta _i}}}\left( n \right)} \right\|}_p}}}} \right) \cdot {\mathcal{{\cal G}}_{{\theta _i}}}\left( n \right),n \in {\left[ {0,1} \right]^{d}}.
\label{generator_individual}
\end{equation} 
The overall framework for generating image-dependent perturbations can be found in Figure~\ref{JUAP_individual}.  
Note that we follow the same  optimization objective and strategy as JUAP in the image-dependent scenario.
We also conduct comprehensive experiments to evaluate the effectiveness of our proposed method for generating image-dependent perturbations in Section \ref{sec:image-dependent}.
It should be noted that JUAP and the related study called \emph{$ADV^{2}$}~\cite{adv2} have different focuses, with JUAP being primarily used for generating \emph{universal adversarial perturbations (UAP)} while \emph{$ADV^{2}$} for \emph{generating image-dependent perturbations (IDP)}. Therefore, we only verify the potential of JUAP in generating IDPs to deceive both classifiers and interpreters without comparing JAP with \emph{$ADV^{2}$}, which is unfair and meaningless.

\begin{algorithm}[t]
  \caption{Our proposed method \textbf{JUAP}}  
  \label{al_juap}
  \KwIn{\\
Benign samples $x$, generator $\mathcal{G_{\theta}}$, classifier $f$, interpreter $\mathcal{I}$, vector norm $p$, threshold $\zeta$, random noise $n$, iteration $T$.\\
\textbf{Output:} Joint universal adversarial perturbation $\hat n$.\\
\textbf{Procedure:}}
  \For{t \text{in} 1:T}
	{
Reconstruct and scale the random noise $n$ to get JUAP $\hat n$:
$\hat n = \min \left( {1,\frac{\zeta }{{{{\left\| {{\mathcal{G}_\theta }\left( n \right)} \right\|}_p}}}} \right) \cdot {\mathcal{G}_\theta }\left( n \right)$ \;
Create adversarial examples $\hat x$:
$\hat x = x + \hat n$\;
Compute the loss function by using one of objectives for attacking classifiers in Eqs.~\eqref{cross}-\eqref{target} \;
Compute the interpretation shift by Eq.~\eqref{foolint} \;
Compute the overall objective $\mathcal{L}_{JUAP}$ by Eq. \eqref{final}\;
Update parameters $\theta$ of the generator: 
$\theta  \leftarrow \theta  - \gamma  \cdot {\nabla _\theta }{\mathcal{L}_{JUAP}}$ \;
	}
\textbf{end for}\\
\end{algorithm}  
\section{Experiments}\label{sec:experiments}
%In this section, we construct comprehensive experiments to demonstrate the effectiveness and robustness of the proposed framework based on three public image classification datasets. 

\subsection{Datasets}
In our experiment, three representative image classification datasets (i.e., ImageNet, Chest X-Ray dataset, and Intel Image Classification dataset) are used to validate the effectiveness of the proposed method. 
The details are shown as follows:
\begin{itemize} %[leftmargin=*] 
    \item \textbf{ImageNet}~\cite{imagenet} is the most widely used image classification dataset, which contains more than 12 million training images, 50,000 validation images, and 100,000 test images with 1,000 classes. We randomly sample 20\% validation images (10,000 images) as training and test sets, respectively.
    \item \textbf{Chest X-Ray Dataset}~\cite{xray} contains 5,863 chest X-ray images from children with two classes (i.e., Pneumonia/Normal). We use the training and test sets to evaluate the performance of the proposed framework.
    \item \textbf{Intel Image Classification Dataset}~\footnote{https://www.kaggle.com/datasets/puneet6060/intel-image-classification} is initially published by Intel to host an Image classification Challenge. It contains around 25,000 images under 6 classes. We randomly choose 20\% training images as the training and test sets, respectively.
\end{itemize}

\subsection{Implementation details}
\subsubsection{\textbf{Classifiers}} 
Two  representative deep neural networks are used as the classifiers, i.e., ResNet18 \cite{resnet} and DenseNet121 \cite{densenet}. 
These two DNNs vary in capacities and architectures, which factors out the influence of different classifiers and demonstrates the robustness of the proposed method. 
We directly use the pre-trained models as the classifiers when constructing experiments on the ImageNet dataset. For other datasets, we fine-tune the pre-trained models as the target classifiers.
The Top-1 accuracy attained on different datasets is summarised in Table~\ref{ori_classify}.

\begin{table}[htbp]
\centering
% \vskip -0.15in
\caption{Top-1 accuracy on different datasets.}%>0.2
% \vskip -0.10in
\label{ori_classify}%
\scalebox{1}
{
\begin{tabular}{|c|c|c|c|c|}
\hline
\multicolumn{2}{|c|}{\textbf{Datasets}}              & \textbf{ImageNet} & \textbf{X-Ray} & \textbf{Intel} \\ \hline \hline
\multirow{2}{*}{\rotatebox{0}{\textbf{DNNs Models}}} & \textbf{ResNet18}     & 67.340  & 87.821  & 90.100          \\ \cline{2-5} 
                                 & \textbf{DenseNet121} & 72.100  & 91.506  & 88.604         \\ \hline
\end{tabular}
}
% \vskip -0.15in

\end{table}

\subsubsection{\textbf{Universal Adversarial Perturbations Generators}}
\label{sec:UAP-G}
In order to evaluate the robustness of our proposed method JUAP, in this paper, we adopt two representative architectures to generate universal adversarial perturbations. 
\begin{itemize} %[leftmargin=*]
    \item \textbf{ResNet generator}~\cite{resnetgenerator} consists of several residual blocks proposed in ~\cite{resnet} for reconstructing the input noise. \emph{Residual connections} make the output image share structure with the input image, which is suitable for the image-to-image translation task.
    \item \textbf{U-Net generator}~\cite{unet} is comprised of five convolutional layers and symmetrical upsampling operators. The skip connections between the convolutional layer and its symmetrical upsampling operator increase the resolution of the output, which is widely used in image translation and semantic segmentation tasks.
\end{itemize}

\subsubsection{\textbf{Baselines}}
To better demonstrate the superiority of the proposed method, we use several representative methods as baselines for comparison.
The details about these baselines and variants of the proposed method are summarised as follows:
\begin{itemize}%[leftmargin=*] 
    \item UAP~\cite{uap}: This is the first iterative method for generating universal adversarial perturbations.
    \item GUAP~\cite{guap}: This method is a typical generative method for creating universal adversarial perturbations.
    \item GUAP-$C_t$~\cite{guap}: This method is used to generate UAPs for the targeted attack.
    \item PGD~\cite{pgd}: This method is used as the representative method to create image-dependent perturbations.
    \item R-JUAP: This is the proposed method that uses the ResNet generator to create JUAPs.
    \item U-JUAP: This method is proposed to create JUAPs based on the U-Net generator.
\end{itemize}

\subsubsection{\textbf{Evaluation Metrics}}
Three evaluation metrics are used to quantitatively evaluate the performance of the proposed method from two aspects, i.e., attacking  DNNs classifiers and interpreters.
The details are summarised as follows:
\begin{itemize} %[leftmargin=*] 
    \item \textbf{Fooling Ratio (\textbf{FR})}~\cite{uap}.
    This metric is used to evaluate the effectiveness of fooling classifiers as follows: 
        \begin{equation}
        % \centerline{
        % $
        \small
        \text{Fooling~Ratio}\left( {\text{FR}} \right) = \left( {\sum\limits_{i = 1}^N {\mathbbm{1}\left( {f\left( {{{\hat x}_i}} \right) \ne f\left( {{x_i}} \right)} \right)} } \right)/N,
        % $
        % }
        \label{fr}
        \end{equation}
    where $\mathbbm{1}(\cdot)$ is the indicator function. The higher FR values indicate better performance for attacking the DNNs classifier.  
    \item \textbf{$\cal{L}_1$ metric (\textbf{L1})}. As for attacking interpreter, this metric is exploited to assess the  discrepancy of attribution maps between benign and adversarial examples, which is computed as follows:
        \begin{equation}
        \label{l1}
        % \centerline{
        {{{\cal L}}_1} = {\left\| {{{\cal I}}\left( {\hat x} \right) - {{\cal I}}\left( x \right)} \right\|_1}.
        % }
        \end{equation}
     The lower $\cal{L}_1$ values indicate better performance for attacking DNNs interpreters.  
    
    \item \textbf{Intersection-Over-Union (\textbf{IOU}) score}. This metric is widely used in the target detection domain. We adopt it to measure the similarity of attribution maps:
        \begin{equation}
        \label{iou}
        %\centerline{
        %$
        \text{IOU}\left( {{{\cal I}}\left( x \right),{{\cal I}}\left( {\hat x} \right)} \right) = \frac{{\left| {O\left( {{{\cal I}}\left( x \right)} \right) \cap O\left( {{{\cal I}}\left( {\hat x} \right)} \right)} \right|}}{{\left| {O\left( {{{\cal I}}\left( x \right)} \right) \cup O\left( {{{\cal I}}\left( {\hat x} \right)} \right)} \right|}},
        %$
        %}
        \end{equation}
where $O(\cdot)$ is the binarization function.
The higher IOU values indicate better performance for attacking interpreters.  

\end{itemize}

\subsubsection{\textbf{Parameter Settings}}
All experiments are implemented on $PyTorch$, and all baselines are conducted by their open-source implementations. For JUAP, we use Adam the optimizer with a learning rate of 0.0002, and the batch size is set to 30. 
All images are resized to 224~$\times$~224 and preprocessed by widely used data augmentation methods, e.g., Random Crop and Random Horizontal Flip. 
During the inference process, all test images are resized to 224~$\times$~224 without further preprocessing.
In Eq.~\eqref{generator}, $p=2$ and $\zeta=2000$ are set as the default. 
In the following experiments, we also set $p=\infty$ and $\zeta=10$ to demonstrate the robustness of the proposed framework. 
The reason for selecting these two sets of parameters is that the generated UAPs can effectively carry out attacks with imperceptible perturbations~\cite{uap, guap}. 
In Eq.~\eqref{final}, $\delta$ is set to -0.8. Due to the differences of characteristics between multiple interpreters, $\lambda$ is fine-tuned dynamically.
We recommend setting $\lambda \in [0.0001, 0.003]$. 
During the training process, we mainly use Eq~\eqref{cmin} as the objective.
When deceiving RTS, since the encoder plays a crucial role in creating attribution maps, solely using the aforementioned loss function $\mathcal{L}_{fool-interpreter}$ is not sufficient to manipulate the interpretation. 
We add an additional loss function to deceive RTS, i.e., ${{{\cal L}}_{rts}} = {\left\| {e\left( {\hat x} \right) - e\left( x \right)} \right\|_2}$, where $e(\cdot)$ is the output of the encoder.

\begin{table}[t]
\centering
% \vskip -0.10in
\caption{Fooling ratio on different datasets.}%>0.2
% \vskip -0.1in
\label{fc1}%
\scalebox{0.9}
{
\begin{tabular}{|c|c|c|c|c|c|}
\hline
\multicolumn{3}{|c|}{\textbf{Datasets}}              & \textbf{ImageNet} & \textbf{X-Ray} & \textbf{Intel} \\
% \multicolumn{2}{|c|}{\textbf{Method}} & \textbf{FR} & \textbf{FR} & \textbf{FR}\\ 
\hline \hline
 \multirow{16}{*}{\rotatebox{90}{\textbf{Method}}} &
\multirow{8}[2]{*}{\textbf{ResNet18}} & \textbf{UAP}   & 77.64  & 75.80  & 51.46  \\
       &   & \textbf{GUAP}  & 79.91  & 83.97  & 50.75  \\
       &   & \textbf{R-JUAP-CAM} & 80.35  & 84.62 & 55.627  \\
       &   & \textbf{R-JUAP-GradCAM} & 80.21  & \textbf{84.94}  & \textbf{51.71} \\
       &   & \textbf{R-JUAP-RTS} & 83.88  & 79.49  & 50.78  \\
       &   & \textbf{U-JUAP-CAM} & 77.35  & 80.29  & 54.63  \\
       &   & \textbf{U-JUAP-GradCAM} & 84.27  & 82.53  & 46.58  \\
       &   & \textbf{U-JUAP-RTS} & \textbf{85.00} & 83.01  & 50.32  \\
    % \hline \hline
    \cline{2-6}
    & \multirow{8}[2]{*}{\textbf{DenseNet121}} & \textbf{UAP}   & \textbf{78.37}  & 75.16  & 46.83  \\
    &      & \textbf{GUAP}  & 77.01  & 79.17  & 53.10  \\
    &      & \textbf{R-JUAP-CAM} & 76.60  & \textbf{83.81}  & 47.54  \\
    &      & \textbf{R-JUAP-GradCAM} & 77.54  & 81.09 & 50.57  \\
    &      & \textbf{R-JUAP-RTS} & 77.31  & 76.12  & 52.99  \\
    &     & \textbf{U-JUAP-CAM} & 77.17  & 79.81  & 45.01 \\
    &      & \textbf{U-JUAP-GradCAM} & 78.10 & 78.53  & 51.78  \\
    &      & \textbf{U-JUAP-RTS} & 76.12  & 79.81  & \textbf{54.06}  \\
    \hline
\end{tabular}
}
% \vskip -0.12in
\end{table}

\begin{table}[t]
\centering
%\vskip -0.15in
\caption{Interpretation discrepancy between adversarial and benign attribution maps (Model: DenseNet121).}%>0.2
% \vskip -0.12in
\label{table_fi1}%
\scalebox{0.95}
{
\begin{tabular}{|c|c|c|c|c|c|c|c|}
\hline
\multicolumn{2}{|c|}{\textbf{Datasets}}              & \multicolumn{2}{|c|}{\textbf{ImageNet}} & \multicolumn{2}{|c|}{\textbf{X-Ray}} & \multicolumn{2}{|c|}{\textbf{Intel}} \\ \hline
\multicolumn{2}{|c|}{\textbf{Metrics}} & \textbf{L1} & \textbf{IOU} & \textbf{L1} & \textbf{IOU} & \textbf{L1} & \textbf{IOU} \\ \hline \hline
    \multirow{12}{*}{\rotatebox{90}{\textbf{Method}}} & \textbf{UAP-CAM} & 0.23  & 0.52  & 0.28  & 0.45  & 0.25  & 0.51  \\
    & \textbf{GUAP-CAM} & 0.18  & 0.59  & 0.25  & 0.46  & 0.20  & 0.53  \\
    & \textbf{U-JUAP-CAM} & 0.17  & 0.63  & 0.22  & 0.53  & 0.18  & 0.59  \\
    & \textbf{R-JUAP-CAM} & \textbf{0.17}  & \textbf{0.63}  & \textbf{0.21}  & \textbf{0.54}  & \textbf{0.18}  & \textbf{0.61}  \\
    \cline{2-8}
    & \textbf{UAP-GradCAM} & 0.21  & 0.36  & 0.28  & 0.44  & 0.23  & 0.55  \\
    & \textbf{GUAP-GradCAM} & 0.18  & 0.46  & 0.25  & 0.47  & 0.20  & 0.53  \\
    & \textbf{U-JUAP-GradCAM} & 0.18  & 0.49  & 0.21  & 0.54  & 0.20  & 0.55  \\
    & \textbf{R-JUAP-GradCAM} & \textbf{0.17} & \textbf{0.51} & \textbf{0.21} & \textbf{0.54} & \textbf{0.20} & \textbf{0.57} \\
    \cline{2-8}
    & \textbf{UAP-RTS} & 0.10  & 0.68  & 0.13  & 0.72  & 0.13  & 0.61  \\
    & \textbf{GUAP-RTS} & 0.14  & 0.60  & 0.14  & 0.71  & 0.20  & 0.54  \\
    & \textbf{U-JUAP-RTS} & 0.08  & 0.72  & 0.08  & 0.81  & 0.12  & 0.64  \\
    & \textbf{R-JUAP-RTS} & \textbf{0.07} & \textbf{0.74} & \textbf{0.06} & \textbf{0.85} & \textbf{0.08} & \textbf{0.71} \\
    \hline
\end{tabular}
}
% \vskip -0.15in
\end{table}

\subsection{Effectiveness of Attacking Classifiers}
We first evaluate the effectiveness of JUAP for attacking target classifiers. We summarise the fooling ratio for different target classifiers on three benchmark datasets, and the results are shown in Table~\ref{fc1}. We only report one result for UAP and GUAP since interpreters do not affect their fooling ratios. We use R-JUAP-CAM (or R-JUAP-GradCAM, and R-JUAP-RTS) and U-JUAP-CAM (or U-JUAP-GradCAM, and U-JUAP-RTS) to represent the JUAP generated by the ResNet generator and U-Net generator, respectively, where the interpreter is CAM (or GradCAM, and RTS). 
From the results, we have the following observations:
\begin{itemize}%[leftmargin=*] 
    \item Compared with other baselines, it is observed that both R-JUAP and U-JUAP achieve a high fooling ratio across different target classifiers, which demonstrates the effectiveness of JUAP for fooling classifiers. 
    Although the optimization objective is complicated, it is still possible to find the optimal solution.
    \item R-JUAP and U-JUAP achieve a similar fooling ratio, which indicates that the proposed framework is robust to the architecture of the generator. 
    Meanwhile, in most cases, R-JUAP and U-JUAP with different interpreters outperform the other baselines in the fooling ratio, which indicates that the proposed framework successfully resolves the dilemma between attacking the classifier and its coupled interpreters.
\end{itemize}

\begin{figure}[t]
% \vskip -0.15in
\centering
\includegraphics[width=1\columnwidth]{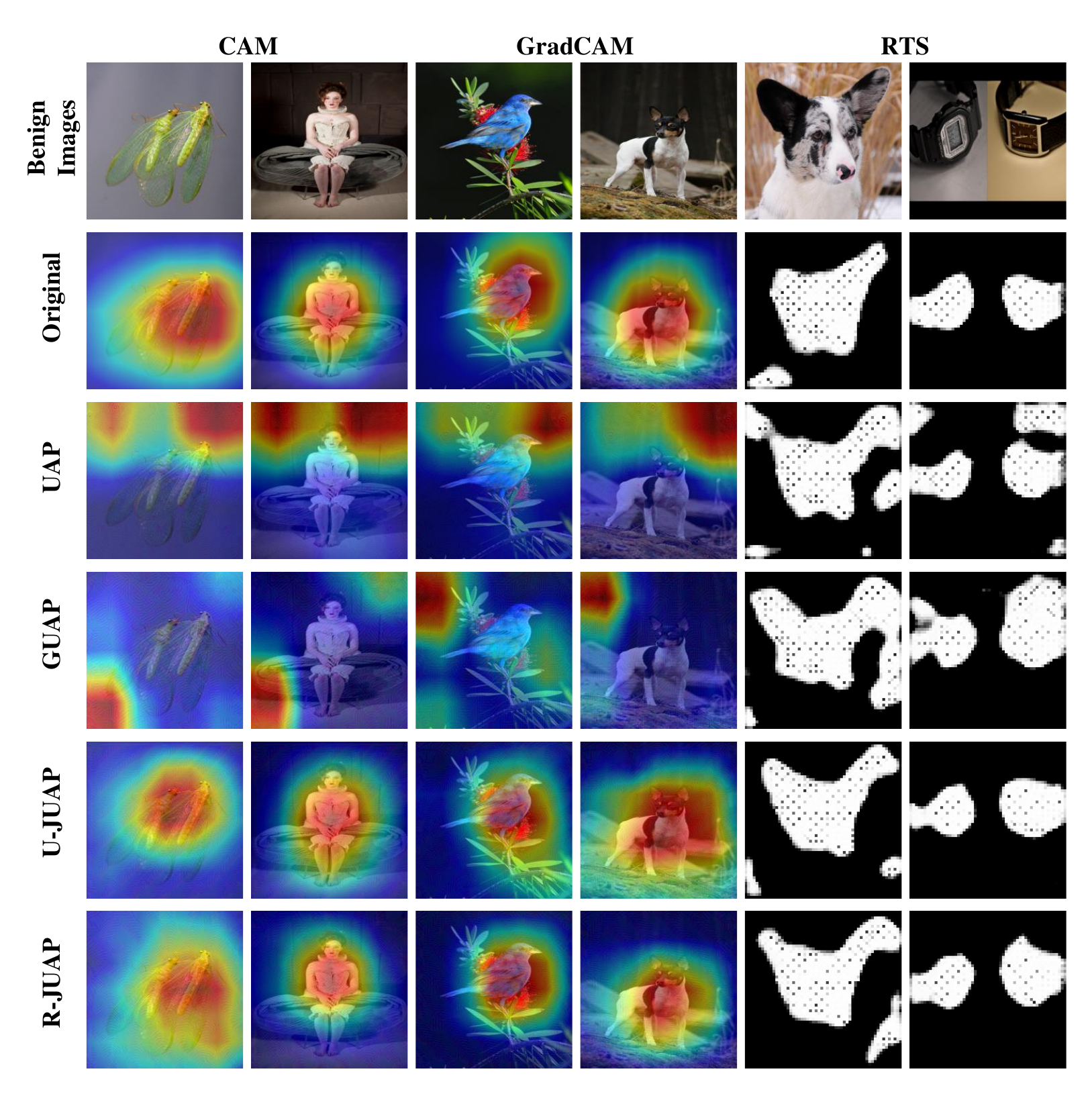}
% \vskip -0.15in
\caption{Attribution maps of benign images and adversarial images with different interpreters on ResNet18.}
\label{sample1}
% \vskip -0.15in
\end{figure}

\subsection{Effectiveness of Attacking Interpreters} \label{exp_ai}
\begin{figure*}[t]
\centering
% \vskip -0.15in
\subfigure[
	\scriptsize
	Interpreter: CAM, Metric: $\cal{L}_1$]{
	\label{fig51}
	\includegraphics[width=1.8in]{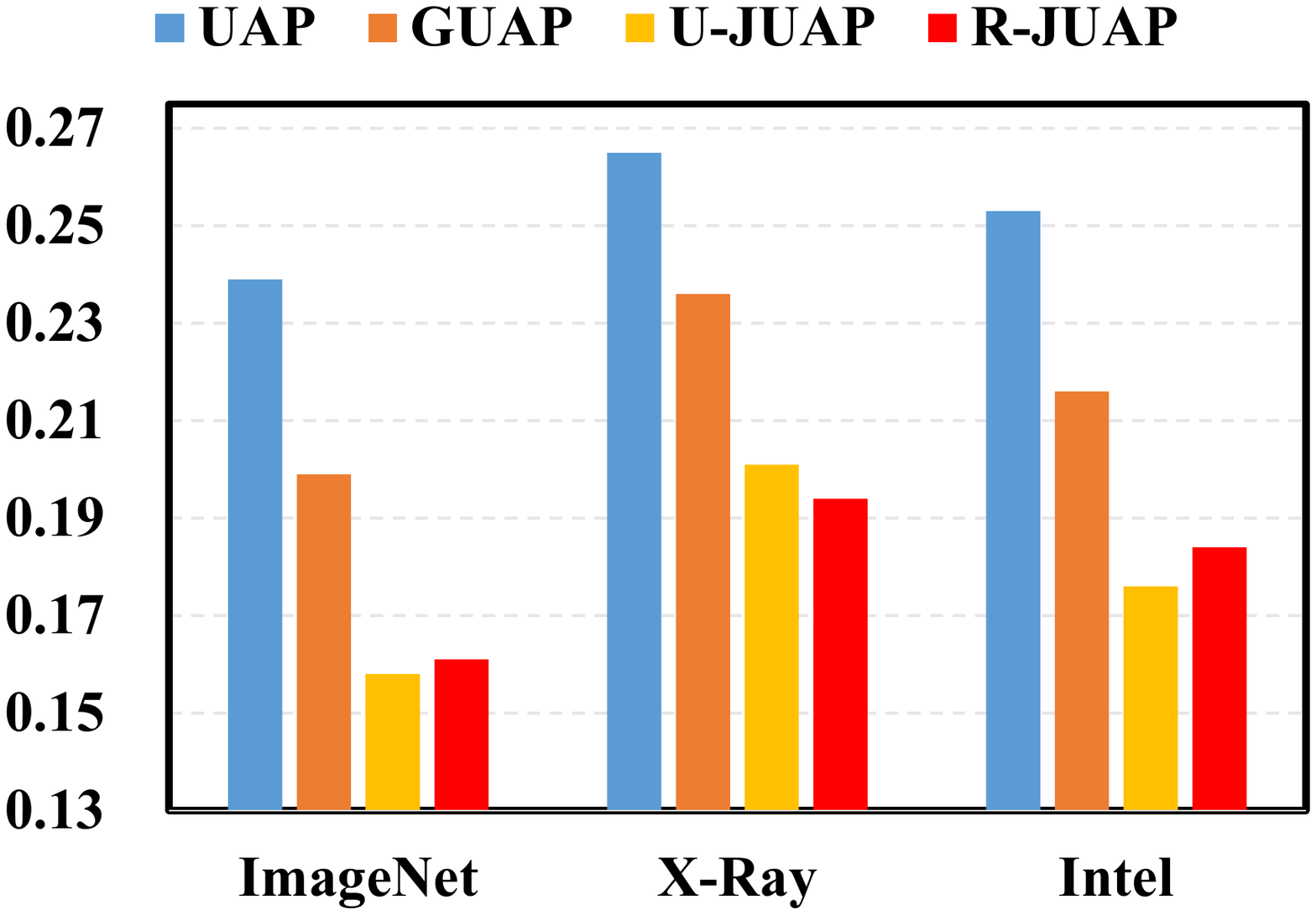}
}
% \vspace{-3mm}
\subfigure[
	\scriptsize
	Interpreter: CAM, Metric: IOU]{
	\label{fig54}
	\includegraphics[width=1.8in]{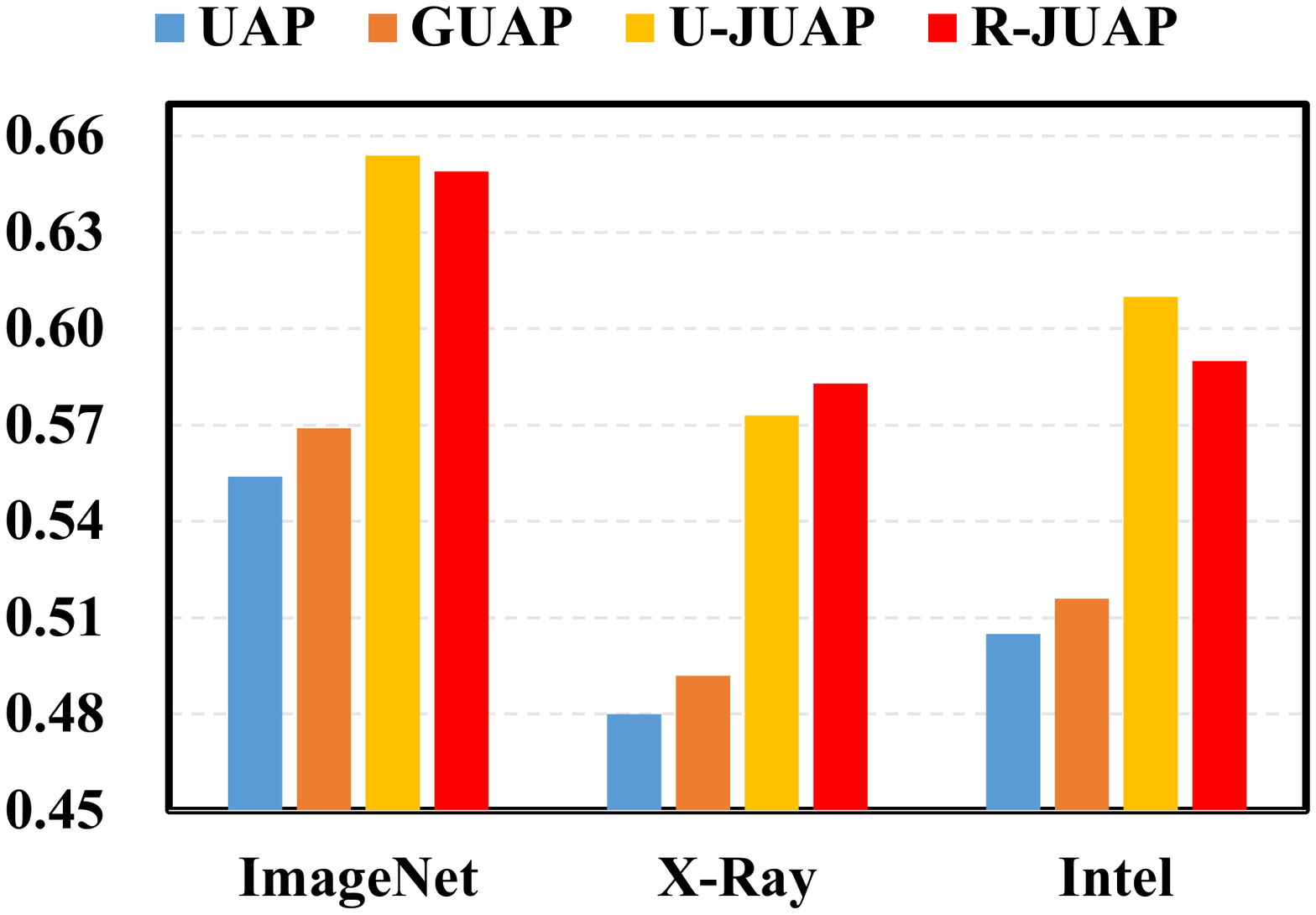}
}
% \vspace{-3mm}
\subfigure[
	\scriptsize
	Interpreter: GradCAM, Metric: $\cal{L}_1$]{
	\label{fig52}
	\includegraphics[width=1.8in]{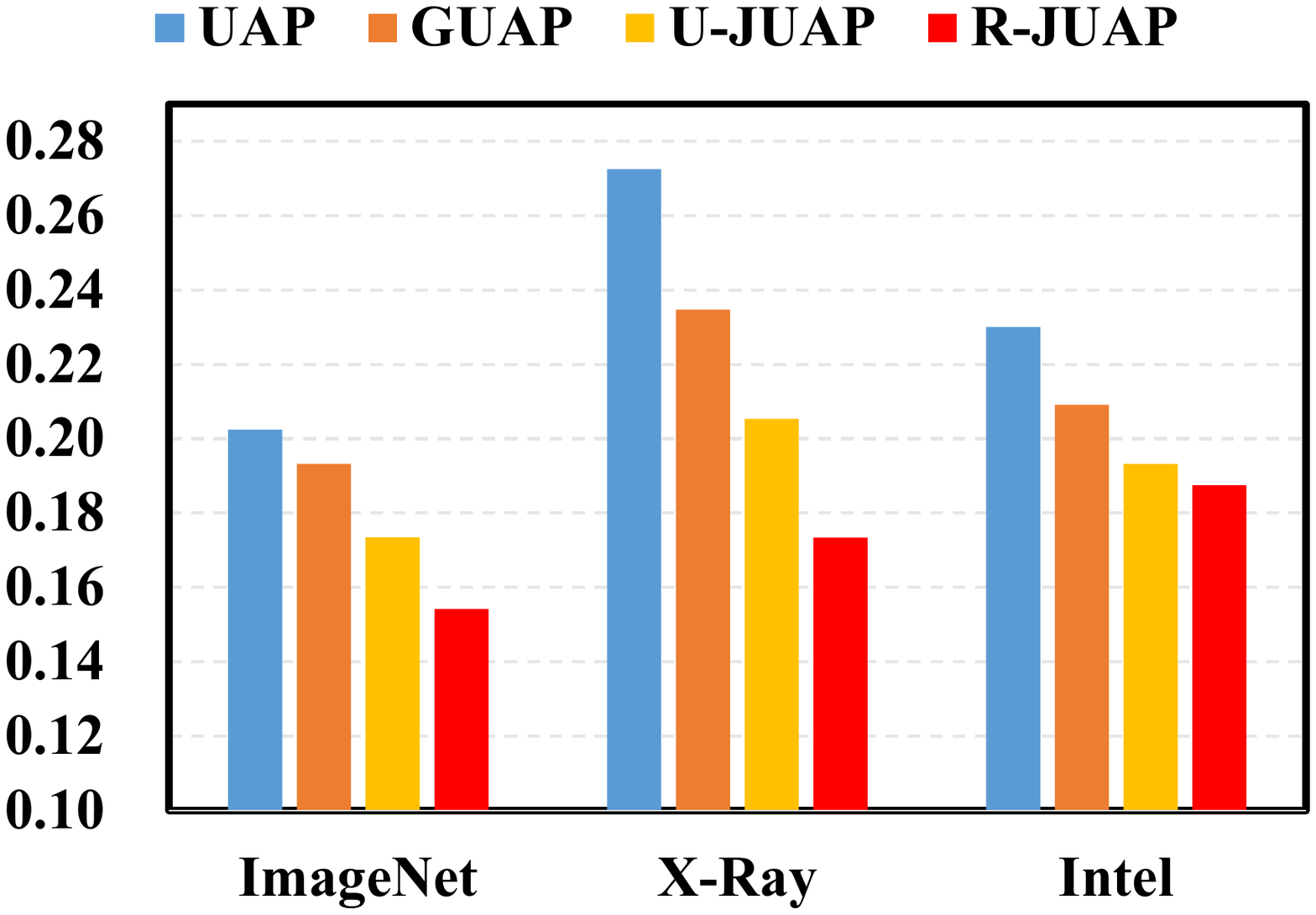}
}
% \vspace{-3mm}
\subfigure[
	\scriptsize
	Interpreter: GradCAM, Metric: IOU]{
	\label{fig55}
	\includegraphics[width=1.8in]{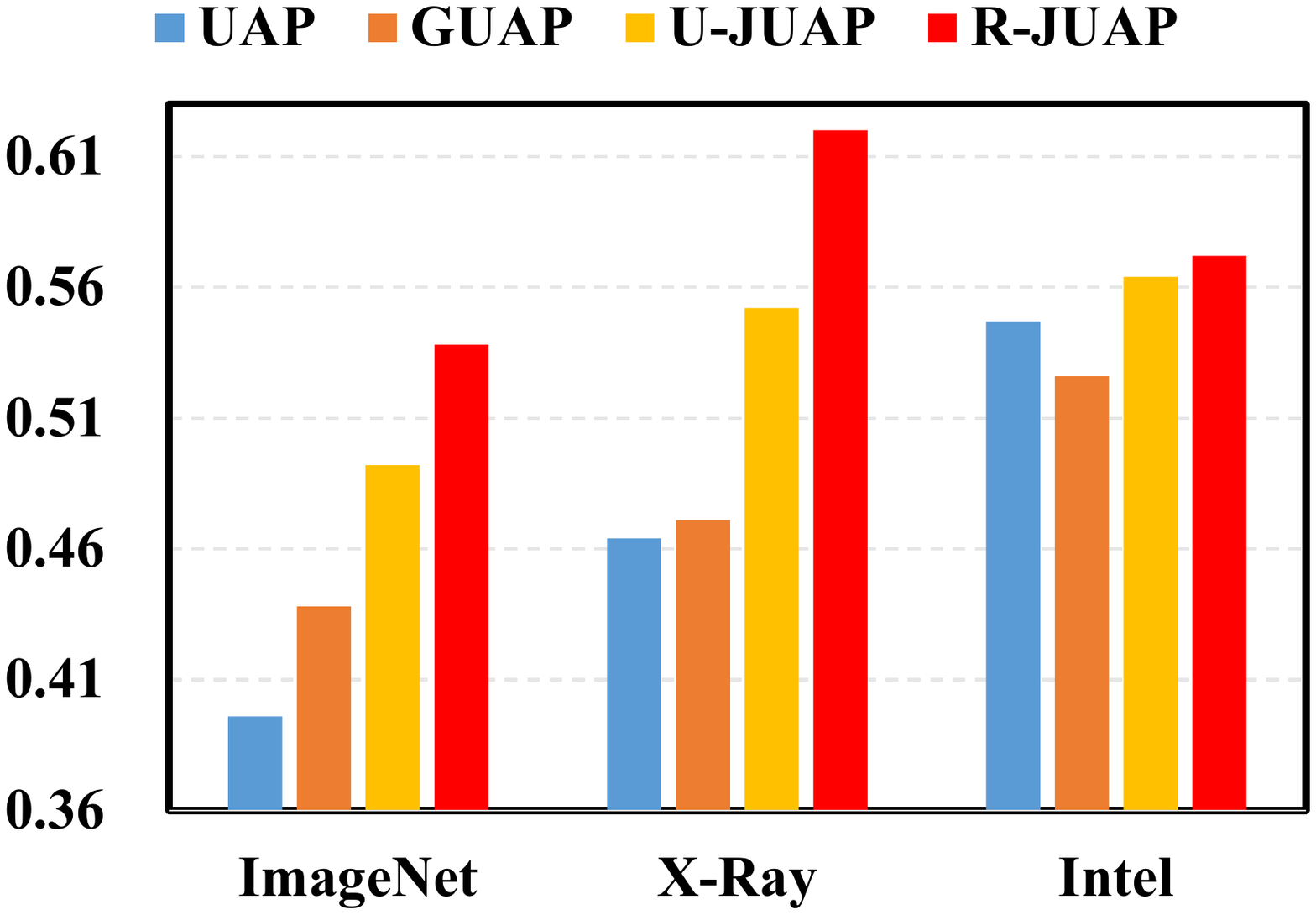}
}
% \vspace{-1mm}
\subfigure[
	\scriptsize
	Interpreter: RTS, Metric: $\cal{L}_1$]{
	\label{fig53}
	\includegraphics[width=1.8in]{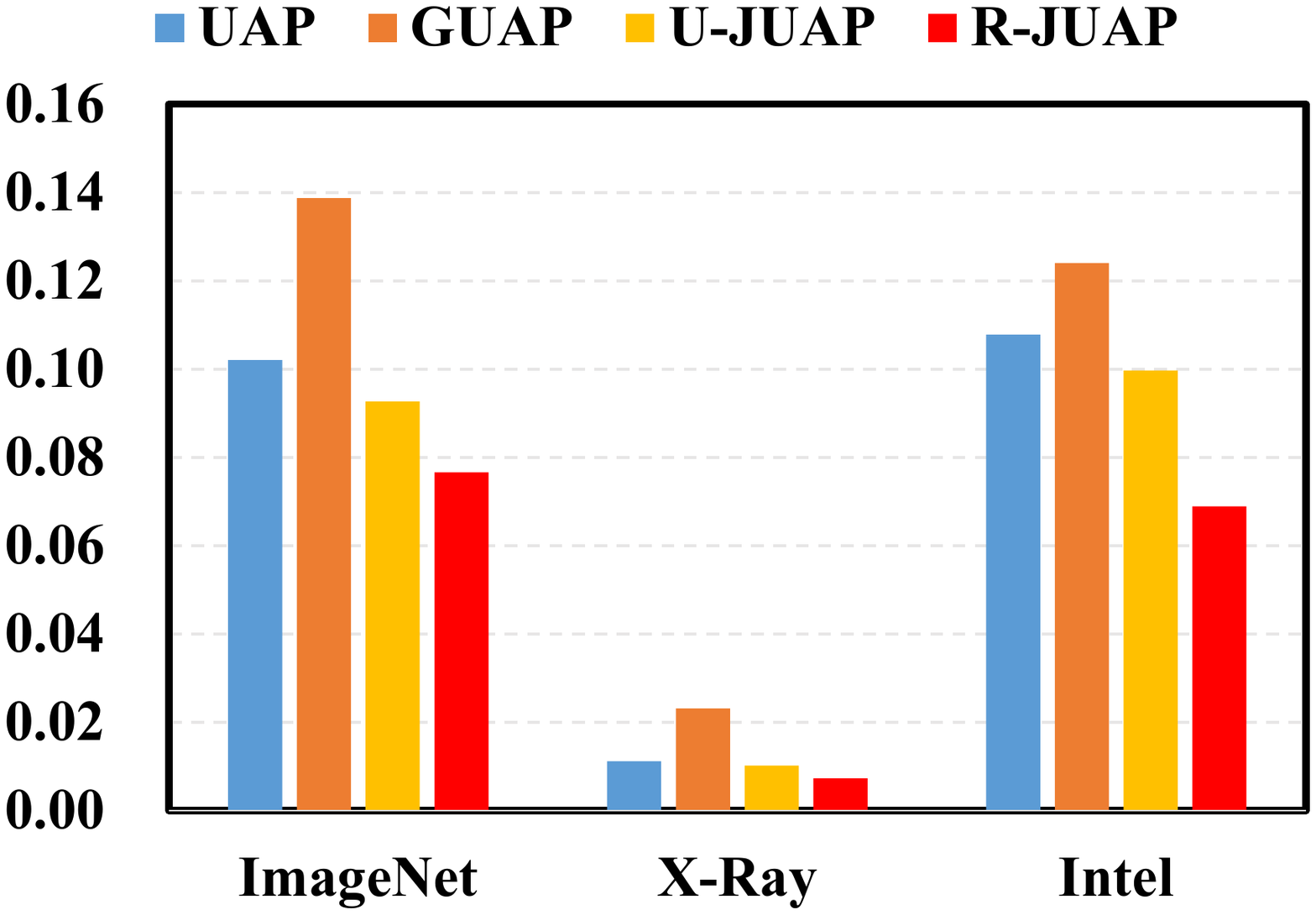}
}
% \vspace{-1mm}
\subfigure[
	\scriptsize
	Interpreter: RTS, Metric: IOU]{
	\label{fig56}
	\includegraphics[width=1.8in]{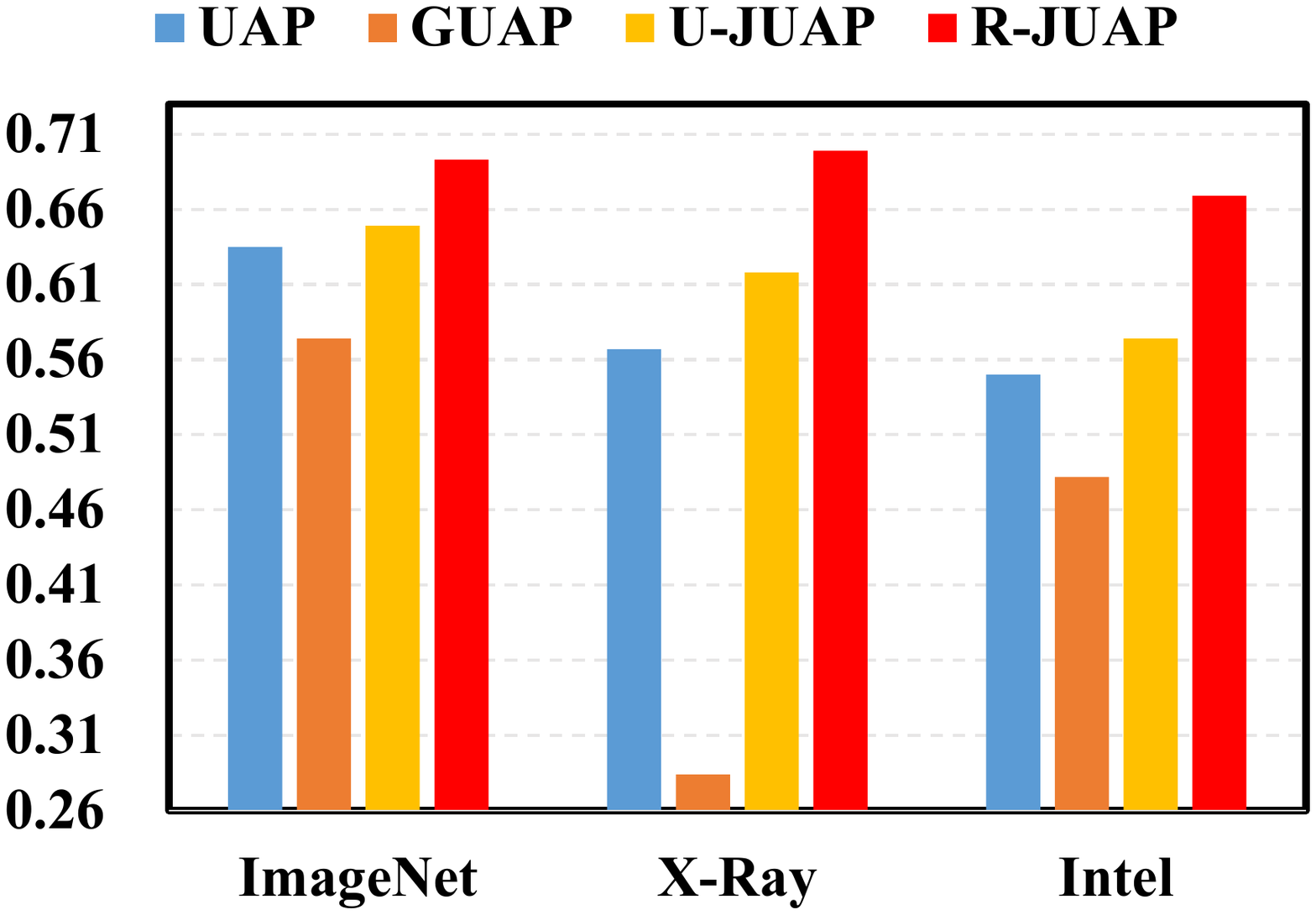}
}
% \vspace{-3mm}
\caption{Interpretation discrepancy of attribution maps between benign and adversarial examples (Model: ResNet18).}
% \vskip -0.15in
\label{image_fi1}
\end{figure*}

In this section, we evaluate the effectiveness of JUAP for deceiving interpreters, i.e., make the attributions maps of benign and adversarial examples similar. We summarise the $\cal{L}_1$ measure and IOU score of adversarial attribution maps with respect to benign maps on three datasets. 
The results are shown in Table~\ref{table_fi1}, Figure~\ref{sample1}, and Figure~\ref{image_fi1}, respectively.
%\wq{Figure~\ref{image_fi1} ???figure 4 5? illustrate the attribution maps between benign images and adversarial images, . }
We get the following insightful observations from the experimental results.

\begin{itemize} %[leftmargin=*] 
    \item We first visualize some examples for qualitative comparison. Figure~\ref{sample1} shows some attributions maps of benign and adversarial images (produced by UAP, GUAP, U-JUAP, and R-JUAP) with respect to different interpreters (i.e., CAM, GradCAM, and RTS). We observe that attribution maps of U-JUAP and R-JUAP are similar with that of the benign images with respect to various interpreters. As for the other baselines, the differences of attribution maps are large, which makes the attack perceptible.
    \item Next we use $\cal{L}_1$ measure and IOU score for quantitative comparison. As shown in Table~\ref{table_fi1} and Figure~\ref{image_fi1}, both R-JUAP and U-JUAP achieve high IOU scores with small $\cal{L}_1$ measure, indicating that the adversarial attribution maps of JUAP are perceptually indistinguishable from their benign counterparts.
    From these comprehensive experiments, we sufficiently demonstrate the effectiveness of JUAP in deceiving both classifiers and interpreters.
\end{itemize}

\begin{figure}[htbp]
\centering
\includegraphics[width=1\columnwidth]{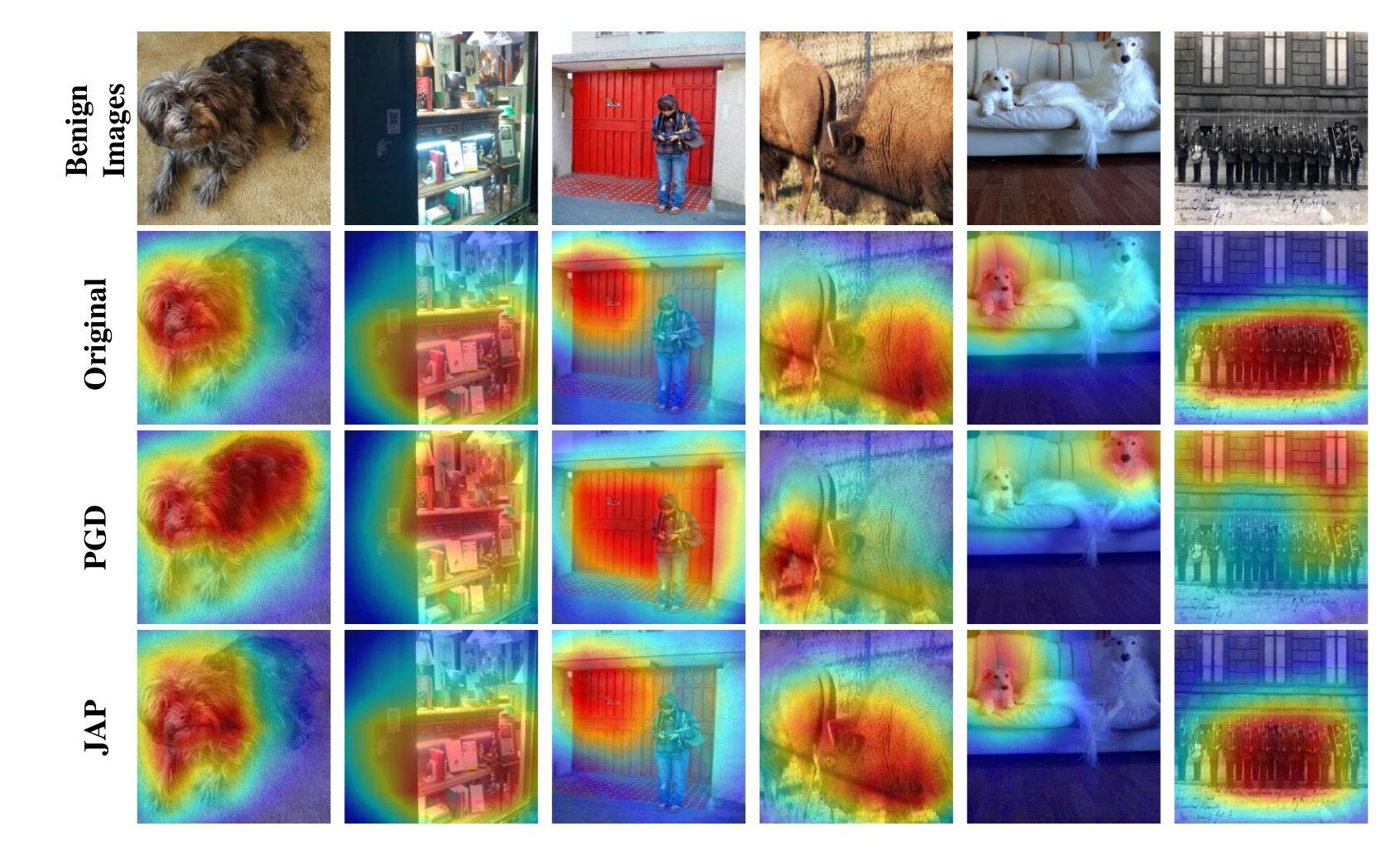}
% \vskip -0.15in
\caption{Attribution maps of benign images and adversarial images based on the GradCAM interpreter (Model: ResNet18).}
\label{sample_pgd}
% \vskip -0.15in
\end{figure}

\subsection{Effectiveness of Generating IDPs}
\label{sec:image-dependent}
As we mentioned before, the proposed framework is equipped with the ability to produce image-dependent perturbations for malicious  attacks. 
In this subsection, we use the most representative attacker, PGD~\cite{pgd}, as the baseline for comparison, and GradCAM is exploited as the basic interpreter. 
We first illustrate some examples for qualitative comparison.
As shown in Figure~\ref{sample_pgd}, attribution maps of samples added joint adversarial perturbations almost remain unchanged. 
The quantitative results are summarized in Table~\ref {individual_attack}. We can observe that JAP achieves the high fooling ratio and IOU with the small $\cal{L}_1$ measure, which further quantitatively demonstrates the above conclusion.

\begin{table}[b]
\centering
% \vskip -0.10in
\caption{Results of PGD and JAP on ImageNet.}%>0.2
% \vskip -0.1in
\label{individual_attack}%
\scalebox{1}
{
\begin{tabular}{|c|c|c|c|c|}
\hline
\textbf{Target Model} & \textbf{Method}             & \textbf{FR} & \textbf{L1} & \textbf{IOU} \\ \hline \hline
    \multirow{2}[0]{*}{\textbf{ResNet18}} & \textbf{PGD}   & 100.00  & 36.70  & 0.55  \\
          & \textbf{JAP}   & 100.00  & \textbf{27.15}  & \textbf{0.65}  \\
    \hline \hline
    \multirow{2}[0]{*}{\textbf{DenseNet121}} & \textbf{PGD}   & 100.00  & 34.54  & 0.57  \\
          & \textbf{JAP}   & 100.00  & \textbf{26.42}  & \textbf{0.66}  \\
    \hline
\end{tabular}
}
% \vskip -0.15in

\end{table}

% \vspace{-3mm}
\subsection{Model Analysis}\label{ra}
In this section, we aim to investigate the impact of model components and model hyper-parameters in our proposed framework. 

\subsubsection{\textbf{$L_p$ Norm Attacks}}
In addition to $p=2$ with threshold $\zeta=2000$ in Eq.~\eqref{generator}, here we investigate how  $p=\infty$ with threshold $\zeta=10$ affects the attacking performance.
The results are summarized in Table~\ref{robutsness_sf}.
We observe that the three metrics (i.e., fooling ratio, $\cal{L}_1$ measure, and IOU score) fluctuate in a small range, which indicates that our JUAP creation is robust under different threat models.

\begin{table}[t]
\centering
% \vskip -0.15in
\caption{Fooling ratio on different datasets.}%>0.2
% \vskip -0.15in
\label{robutsness_sf}%
\scalebox{0.85}
{
\begin{tabular}{|c|c|c|c|c|c|c|c|}
\hline
\multicolumn{2}{|c|}{\textbf{Models}}              & \multicolumn{3}{|c|}{\textbf{ResNet18}} & \multicolumn{3}{|c|}{\textbf{DenseNet121}} \\ \hline
\multicolumn{2}{|c|}{\textbf{Method}} & \textbf{FR} & \textbf{L1} & \textbf{IOU} & \textbf{FR} & \textbf{L1} & \textbf{IOU}\\ \hline \hline
    \multirow{6}[0]{*}{\textbf{$ L_{\infty}$-10}} & \textbf{R-JUAP-CAM} & 85.32  & 0.12  & 0.64  & 75.50  & 0.13  & 0.61  \\
          & \textbf{R-JUAP-GradCAM} & 78.39  & 0.15  & 0.66  & 78.88  & 0.17  & 0.61  \\
          & \textbf{R-JUAP-RTS} & 80.38  & \textbf{0.08}  & 0.\textbf{69}  & 76.69  & \textbf{0.07}  & \textbf{0.73}  \\
          & \textbf{U-JUAP-CAM} & 81.69  & 0.16  & 0.64  & 78.37  & 0.17  & 0.62  \\
          & \textbf{U-JUAP-GradCAM} & \textbf{86.30}  & 0.18  & 0.49  & \textbf{79.15}  & 0.18  & 0.47  \\
          & \textbf{U-JUAP-RTS} & 85.44  & 0.09  & 0.64  & 79.06  & 0.09  & 0.68  \\
      \hline
    \multirow{6}[0]{*}{\textbf{$ L_2$-2000}} & \textbf{R-JUAP-CAM} & 80.35  & 0.16  & 0.65  & 76.60  & 0.17  & 0.63  \\
          & \textbf{R-JUAP-GradCAM} & 80.21  & 0.15  & 0.54  & 77.54  & 0.17  & 0.51  \\
          & \textbf{R-JUAP-RTS} & 83.88  & \textbf{0.08}  & \textbf{0.69}  & 77.31  & \textbf{0.07}  & \textbf{0.74}  \\
          & \textbf{U-JUAP-CAM} & 77.35  & 0.15  & 0.65  & 77.17  & 0.17  & 0.63  \\
          & \textbf{U-JUAP-GradCAM} & 84.27  & 0.17  & 0.49  & \textbf{78.10}  & 0.18  & 0.49  \\
          & \textbf{U-JUAP-RTS} & \textbf{85.00}  & 0.09  & 0.65  & 76.12  & 0.08  & 0.72  \\
    \hline
\end{tabular}

}
% \vskip -0.15in
\end{table}

\begin{table}[htbp]
\centering
% \vskip -0.10in
\caption{Results of JUAP with different loss functions.}%>0.2
% \vskip -0.15in
\label{robutsness_lf}%
\scalebox{1}
{
\begin{tabular}{|c|c|c|c|c|}
\hline
\multicolumn{2}{|c|}{\textbf{Method}}              & \textbf{FR} & \textbf{L1} & \textbf{IOU} \\ \hline \hline
    \multirow{2}[0]{*}{\textbf{Non-targeted}} & \textbf{R-JUAP-$ C_{min}$} & 80.35  & 0.16  & 0.65  \\
          & \textbf{R-JUAP-CE} & \textbf{89.63}  & 0.16  & \textbf{0.65}  \\
    \hline
    \multirow{2}[0]{*}{\textbf{Targeted}} & \textbf{GUAP-$ C_t$} & 73.44  & 0.24  & 0.49  \\
          & \textbf{R-JUAP-$ C_t$} & \textbf{77.46}  & \textbf{0.22}  & \textbf{0.53}  \\
    \hline
\end{tabular}
}
% \vskip -0.2in
\end{table}

\subsubsection{\textbf{Objective on Attacking DNNs Classifier}}
In Section \ref{methodology_JUAP}, we introduce three kinds of objectives to optimize the parameters of the generator. 
Here, we study the impact of these different objectives on the proposed framework. 
The coupled interpreter is CAM. 
We use R-JUAP-CE, R-JUAP-$C_{min}$, and R-JUAP-$C_t$ to denote the JUAP optimized by Eqs.~\eqref{cross}-\eqref{target}, respectively. The results are shown in Table~\ref{robutsness_lf}. 
For the non-targeted attack, we can see that using Eq. ~\eqref{cross} and Eq. ~\eqref{cmin} as optimization objectives can achieve similar results. 
For the targeted attack, GUAP-$C_t$ is used as the baseline since it can also produce targeted universal perturbations. As shown in Table~\ref{robutsness_lf}, R-JUAP-$C_t$ outperforms GUAP in $\cal{L}_1$ and IOU scores with a similar fooling ratio.

\section{Related works}\label{sec:rw}
In this section, we briefly review some related work about adversarial attacks and interpretation models.

\subsection{Universal (Image-agnostic) Adversarial Perturbations}
Deep neural networks are susceptible to deception by adversarial attacks. That is, the output of a DNNs model can be fooled by adding well-designed, imperceptible perturbations to the input \cite{fgsm, pgd}. 
More specifically, by considering image-agnostic perturbations,  adversaries can find universal adversarial perturbations (UAPs) that can deceive the neural network model when added to most images. 
Since only one perturbation is required to be generated, UAP has a significant advantage in computational efficiency over image-dependent perturbations \cite{fastuap}. 
Moosavi-Dezfooli et al. \cite{uap} presented the first data-driven iterative UAP generation algorithm that uses \emph{Deep Fool} \cite{deepfool} for each unsuccessfully attacked image to identify the minimum perturbation that can deceive the model and add it to the UAP.
Although the iterative approach can generate UAPs with a high attack success rate, its computational cost is quite expensive. Therefore, generative methods are applied to generate UAPs~\cite{guap,psgan, nag}. Hayes et al. \cite{hayes2018learning} suggested using a universal adversarial network (UAN) to learn the overall distribution of UAPs instead of fixing one UAP. Thus the UAN can quickly generate different UAPs depending on the different input noises. Mopuri et al. \cite{nag} further introduced a new loss function to ensure that the generated perturbations are diverse enough and can fool the neural network model successfully.

\subsection{DNNs Interpretability} 
Interpreters aim to reveal the underlying reasons why a deep neural network model makes a particular prediction in a human-understandable way. 
Several interpretation methods have been proposed for analyzing the discriminative regions of image classification models \cite{grad, gradcam, wgradcam}. 
\emph{CAM} \cite{cam} is the first proposed activation-based interpretation method, which performs a weighted sum of the feature maps of the convolutional layers, usually the last layer, for the explanation. 
To overcome the drawback that \emph{CAM} needs to modify the network architecture, \emph{Grad-CAM} \cite{gradcam} suggested taking the mean value of the gradient as the weight of the feature map and keeping only the regions that have a positive impact on the classification.
\emph{Grad-CAM++} \cite{grad-cam++} weights the gradients for the problem that there may be more than one class in the image, making the localization more accurate. 
Another category of interpretation methods are gradient-based methods, which calculate the gradient of the output over the input image to determine which pixels have a stronger impact on the prediction \cite{Smoothgrad}. 
However, the saliency maps obtained by such methods are usually not clear enough. Instead of relying on gradients, the model-based interpreters train a masking model to predict saliency maps during forward propagation~\cite{rts}.

However, researchers have found that these interpreters can be manipulated arbitrarily. 
For instance, Heo et al. \cite{heo2019fooling} fine-tuned the neural network model by adding the interpretation results to the penalty term of the loss function, which allowed manipulating the saliency map of the interpreter without compromising the accuracy of the model. 
Ghorbani et al. \cite{ghorbani2019interpretation} introduced adversarial perturbations to interpreters, i.e., it is possible to generate interpretations that differ significantly from the original image by simply adding perturbations to the image that are imperceptible to the naked eye and that do not change the classification result. 
Dombrowski et al. \cite{dombrowski2019explanations} further manipulated the interpretation methods by using adversarial perturbations to generate arbitrary explanation maps and explained this phenomenon in terms of the geometric properties of neural networks. 

\subsection{Fooling Both Classifiers and Interpreters} 
Traditional adversarial attack methods change the explanation maps of neural network models while deceiving their predictions, and thus can be easily detected by users. 
A more threatening class of adversarial attacks is one that can mislead neural network models without changing the explanation maps.
By considering \emph{image-dependent perturbations}, Zhang et al. \cite{adv2} proposed an attack method, \emph{$ADV^{2}$}, to integrate the difference between the attribution maps of the generated adversarial examples and the original images into the loss function to ensure that the explanation maps are not changed when misleading the classifier.
Zhang et al. also pointed out that a potential reason why classifiers and interpreters can be successfully attacked simultaneously is that interpreters usually explain only part of the classifier's behavior. Adversarial patches are a more practical class of adversarial attacks that can effectively mislead classifiers, but interpretation methods such as \emph{Grad-CAM} \cite{gradcam} can highlight patch locations and thus be noticed. Subramanya et al. \cite{subramanya2019fooling} optimize the patches while suppressing the heat map of the patch locations, making the adversarial patches unnoticeable to the interpreter when deceiving the classifier. 
However, most of these existing attacking methods focus on image-dependent attacks, i.e., they need to generate specific perturbations for each image.
To the best of our knowledge, this is the first effort to investigate \textbf{universal adversarial perturbations} to jointly deceive the DNNs classifier and interpreter for a trustworthy DNNs model.

\section{Conclusion}\label{sec:conclusion}

In this work, we first find that DNNs models’ interpretations can be used to inspect the problematic outputs from adversarial examples caused by universal adversarial perturbations. 
This finding motivates us to further raise a new research problem: \emph{whether there exist universal adversarial perturbations that can perform a joint attack to attack DNNs classifier and its interpretation simultaneously with malicious desires?}
To address this problem, we propose a generative framework to create the joint universal adversarial perturbations to jointly attack the DNNs classifier and its coupled interpreters. 
Our comprehensive experiments validate the effectiveness of the proposed JUAP  on three real-world datasets. 
Considering the vulnerability of DNNs interpreters, we would like to investigate a new interpretation method to defend against adversarial attacks for trustworthy DNNs in the future.

% \newpage 
\balance
\bibliographystyle{IEEEtran}
% \large{
% \bibliography{references/references}
%  }
\bibliography{Main}

% \newpage

 

% \end{IEEEbiography}



% \vfill

\end{document}